\documentclass[12pt]{article}
 \usepackage{amssymb}
 \usepackage{amsmath}
 \usepackage[utf8]{inputenc}
 \usepackage[T1]{fontenc}
\usepackage{times}
\usepackage{multirow}
\usepackage{hyperref} 

\date{}

 \usepackage{graphicx}
 \usepackage{epsfig}

\newtheorem{theorem}{Theorem}[section]

\newtheorem{e-proposition}[theorem]{Proposition}

\newtheorem{e-definition}[theorem]{Definition\rm}

\newtheorem{theoreme}{Th\'eor\`eme}[section]

\newtheorem{proposition}[theoreme]{Proposition}

\begin{document}
 \begin{center}
{\LARGE {Uniform weak convergence of poverty measures with relative poverty lines}}
 \end{center}
 \begin{center}
 Cheikh Tidiane Seck\\
 {\small \it Département de Mathématiques, UFR SATIC, Universit\'e Alioune Diop, Bambey, S\'en\'egal.}\\
 and\\
  Gane Samb Lo\\
 {\small \it Département de Mathématiques, UFR SAT, Universit\'e Gaston Berger, Saint-Louis, S\'en\'egal, Universit\'e Paris 6, France.}\\
\end{center}

\begin{center}\textbf{Abstract} \end{center}
\indent This paper introduces a general continuous form of poverty index that encompasses  most of the existing formulas in the literature. We then propose a consistent estimator for this index in case the poverty line is a functional of the distribution. We  also establish a uniform functional Central Limit Theorem for the proposed estimator over a suitable product class of real-valued functions. As a consequence,  testing procedures based either on  single or  simultaneously on several poverty indices can be  developed. A simulation study showing the asymptotic normality of the estimator is given as well as an application to real data for estimating the effect of  relative poverty lines on the variance of  the poverty estimates. \\

 
 \textit{\textbf{2010 AMS  Classification}} : 60F12, 62G10, 62G20, 62G30, 62P20.

\section{Introduction}
Let $Y$ be a positive random variable representing the income (or expenditure) distribution in a given population. Suppose that $Y$ is defined on a probability espace $(\Omega,\mathcal{A},\mathbb{P})$, with underlying continuous distribution function $G(y)=\mathbb{P}(Y\leq y),\;\forall y\geq 0.$  Given a   poverty line $z>0$, we say that an individual is poor if his income  is less than $z$. Most of  the poverty measures employed in practice may be represented, in their continuous form, by the following functional :
 \begin{equation} \label{s1}
J(w,f)=\int^{z}_{0}w[G(y),G(z)]f(y,z)dG(y),
\end{equation}
where $w(u,v)$ is a function of $(u,v)\in [0,1]^2$, which is decreasing with respect to its first argument $u$. It is interpreted as a weighting function associated with the kind of poverty measure being considered ; while $f(y,z)$ is called the \textit{poverty deprivation function}, which evaluates the individual contribution of each poor to the global poverty within the population. Note that the function $f(y,z)$ is also decreasing with respect to its first argument $y$.\\

 \indent Formula (\ref{s1}) is quite general and represents a wide class of poverty indices including the most currently used in practice. For instance, the additively decomposable family of poverty measures can be obtained from this formula \eqref{s1} by letting the weighting function $w(u,v)\equiv 1$, for all $(u,v)\in \left[0,1\right]^2$. As well, the non-additively decomposable poverty measures  such as the Sen-like poverty indices may  also be put in the form (\ref{s1}), with specific weighting  and deprivation functions $w(\cdot,\cdot)$ and $f(\cdot,\cdot)$. In the Table \ref{tab0} below, we give some examples of poverty indices with their own weighting and deprivation functions.
 \begin{table}[htbp]
\begin{center}
$$\begin{array}{ccc}
\hline
 \text{Poverty index} &  w(u,v) &  f(y,z)\\
 \hline
\text{FGT}(\alpha)  &         1  &    (1-y/z)^{\alpha}  \\
\text{Sen}  &      2(1- u/v)     &    (1-y/z)   \\
\text{Shorrocks}  & 2(1-u)     & (1-y/z)       \\
\text{Kakwani} (k)     &  (k+1)(1-u/v)^k    &  (1-y/z)^{k}     \\
\text{Watts}    &      1    & \log (z/y)     \\
\hline
\end{array}
$$
\end{center}
\caption{Examples of poverty indices with their weighting and deprivation functions.}
\label{tab0}
\end{table}

Our main goal in this paper is to propose an estimator for the theoretical functional $J(\cdot,\cdot)$, and study its asymptotic behavior by considering a relative poverty line. Indeed, empirical studies with fixed poverty lines are plentiful in the literature ( see, e.g.,\cite{kak1}, \cite{bfz},\cite{rg}, \cite{lss}). But most of them do not take account of the sampling error of the poverty line, which may increase or reduce the precision of the poverty estimates (see, e.g. \cite{p}). Investigating relative poverty in US, Zheng \cite{zh2}  proposed an approach which consider the poverty line as a percentage of the mean or a quantile of the distribution function. He dealt with additively separable (or decomposable) poverty measures, and found that the sampling error associated with poverty lines increases the standard error of the poverty estimates.\\

\indent In this paper, we propose an inference procedure which somewhat extends Zheng's \cite{zh2}  approach to  non-additively decomposable poverty measures including  Sen, Shorrocks  and Kakwani indices as well as many other poverty measures available in the literature. Note that the Kakwani's family  is the most interesting family of poverty indices, because it contains the FGT (Foster-Greer-Thorbecke) and Sen measures, and satisfies most of the normative  axioms desirable on on poverty index. Our methodology is inspired by the modern empirical process theory developed in van der Vaart and Wellner \cite{vdv}, which permits us to obtain the uniform weak convergence of a wide class of empirical  poverty estimators.\\ 

\indent The rest of the paper is organized as follows. In Section 2, we construct an estimator for the functional $J(w,f)$, and establish its asymptotic properties. In Section 3, we derive  testing procedures which allows  to make comparisons based either on one single poverty measure or simultaneously several poverty indices. Section 4 shows, in a simulation study,  the asymptotic normality of the proposed estimator. In Section 5, we give an application to real data sets to estimate the effect of the relative poverty line on the variance of the  some particular poverty estimates. Finally, we prove the main theorem in Appendix.

\section{Construction of the estimator and asymptotic results}
 Let  $Y_1,\cdots, Y_n$ be an independent and identically random sample of the income variable $Y$, with density probability function $g$. Whenever the poverty line $z>0$ is fixed, a direct estimator of \eqref{s1} can be defined as follows :
\begin{equation} \label{s2}
J_n(w ,f) =\frac{1}{n}\sum_{j=1}^{n}
w[G_n(Y_{j,n}),G_n(z)]f(Y_{j,n},z) \mathbb{I}(Y_{j,n}\leq
z),\qquad 
\end{equation}
for any specific functions $w$ and  $f$. Here $\mathbb{I}(\cdot)$ designs the indicator function,  $Y_{1,n}\leq\cdots\leq Y_{n,n}$ are  order statistics associated with the sample  $Y_1,\cdots, Y_n$  and $G_n(y)=\frac{1}{n}\sum_{j=1}^{n}\mathbb{I}(Y_{j}\leq y)$ is the  corresponding empirical distribution function.\\
 In contrast,  when we are concerned with relative poverty analysis, the poverty line becomes a functional of the distribution  $G$, say $z=z(G)$, and then must be estimated. Let $\hat{z}$ be a consistent estimator of $z$ such that the following representation $(R)$  (see, e.g. Thuysbaert and Zitikis \cite{tz}) :
$$ (R) \qquad \hat{z}= z + \frac{1}{n}\sum_{j=1}^{n}\zeta(Y_j)+ o_{\mathbb{P}}(n^{-1/2}), $$
where $\zeta(\cdot)$ is a real-valued function such that ${\rm Var}[\zeta(Y)]=\int_0^{\infty} \zeta^2(y)dG(y)<\infty$. \\
The function $\zeta(\cdot)$ may be equal to the constant 0, in which case $\hat{z}=z$ corresponds to an absolute poverty line. If the poverty line is set to a fraction $k$ of the mean of the distribution, i.e. ${z}=k\mu_G$, with $\mu_G=\int ydG(y)$, then $\zeta(y)=ky$. Whenever a fraction $k$ of a $p$-quantile is chosen, i.e. ${z}=kG^{-1}(p)$,  the Bahadur representation yields $\zeta(y)=\frac{k}{g(G^{-1}(p))}\mathbb{I}(y\leq G^{-1}(p) ).$\\
 Substituting $z$ for $\hat{z}$ in \eqref{s2}, we obtain a novel estimator 
\begin{equation} \label{s3}
\hat{J}_n(w ,f) =\frac{1}{n}\sum_{j=1}^{n}
w[G_n(Y_{j,n}),G_n(\hat{z})]f(Y_{j,n},\hat{z}) \mathbb{I}(Y_{j,n}\leq
\hat{z}).
\end{equation}
Now, we have to prove that $\hat{J}_n(w,f)$ converges almost surely to $J(w,f)$ for any specific functions $w$ and $f$. Because of certain normative properties desirable on a poverty index, the  functions $w$ and $f$ should satisfy some regularity conditions ; that is they belong to the following classes of functions $\mathcal{W}$ and $\mathcal{F}$, respectively :
\begin{eqnarray*}
\mathcal{W}&=&\{w:[0,1] \times [0,1] \rightarrow \mathbb{R}_{+}, \; w
\;\;\text{  continuous,}\;\;\text{and}\;\;u\mapsto w(u,\cdot)
\;\;\text{is non-increasing}\}\\
\mathcal{F}&=&\{f:\mathbb{R}_{+}\times\mathbb{R}_{+}\rightarrow \mathbb{R}_{+},\;f
\;\;\text{ continuous,}\;\;\text{and}\;\;y\mapsto f(y,\cdot)
\;\;\text{is non-increasing}\}.
\end{eqnarray*}
We also need the following conditions (A.1-2) to establish our asymptotic results :
\begin{itemize}
\item[(A.1)] The functions $(u,v)\mapsto w(u,v)$ and $(y,z)\mapsto f(y,z)$ are differentiable  with continuous first-order partial derivatives ;
\item[(A.2)] $\mathcal{W}$ and $\mathcal{F}$ are both pointwise measurable classes of functions. That is, they contain each one a countable subclass $\mathcal{G}$ such that for all $\phi\in\mathcal{G}$, there exists a sequence $\{\phi_m\}_{m\geq 1} \subset \mathcal{G}$, \noindent with $\phi_m(y)\rightarrow \phi(y)$ for every $y$.\\
\end{itemize}
\indent It is established in \cite{ls} that the estimator $J_n(w,f)$, with $z$ fixed, converges almost surely to $J(w,f)$ for any given functions $w\in \mathcal{W}$ and $f\in\mathcal{F}$. 
 In the following proposition, we gives the almost sure consistency of  the estimator $\hat{J}_n(w,f)$ for $J(w,f)$. 
\begin{proposition}\label{l1}
 For any couple of functions $(w,f)\in\mathcal{W}\times\mathcal{F}$, one has with probability 1,
 $$ \hat{J}_n(w,f)\longrightarrow J(w,f),\qquad n\longrightarrow\infty.$$
 \end{proposition}
 \textbf{Proof}. It suffices to show that for all $(w,f)\in\mathcal{F\times W}$, $\hat{J}_n(w,f)$ is asymptotically equivalent to ${J}_n(w,f)$ plus an additional quantity of the form $a(\hat{z}-z)$, where the factor $a$ will be specified later on. Let's decompose $\hat{J}_n(w,f)$ as follows :
 \begin{eqnarray*}
 \hat{J}_n(w,f) &= &\frac{1}{n}\sum_{j=1}^{n}w[G_n(Y_{j,n}),G_n({z})]f(Y_{j,n},{z}) \mathbb{I}(Y_{j,n}\leq {z})\\
 &= &\frac{1}{n}\sum_{j=1}^{n} \left\{ w[G_n(Y_{j,n}),G_n(\hat{z})]f(Y_{j,n},\hat{z})- w[G_n(Y_{j,n}),G_n({z})]f(Y_{j,n},{z})\right\} \mathbb{I}(Y_{j,n}\leq {z})\\
 &=& \frac{1}{n}\sum_{j=1}^{n} w[G_n(Y_{j,n}),G_n(\hat{z})]f(Y_{j,n},\hat{z})\left\{\mathbb{I}(Y_{j,n}\leq \hat{z})-\mathbb{I}(Y_{j,n}\leq {z}) \right\} \\
 &=:& I+II+III
 \end{eqnarray*}
One can readily observe that the first term $I$ is exactly $J_n(w,f)$. \\
\indent By applying the mean value theorem, the second term $II$ becomes
\begin{eqnarray*}
II&=&\frac{1}{n}\sum_{j=1}^{n}\left\{ \frac{\partial}{\partial v} w[G_n(Y_{j,n}),G_n(z_0)]f(Y_{j,n},\hat{z})[G_n(\hat{z})-G_n(z)] \right\}\mathbb{I}(Y_{j,n}\leq {z})\\
&+ &  \sum_{j=1}^{n}\left\{\frac{\partial}{\partial z}f(Y_{j,n},z_0)w[G_n(Y_{j,n}),G_n(z)][\hat{z}-z]\right\}\mathbb{I}(Y_{j,n}\leq {z}),
\end{eqnarray*}
where $z_0$ is between $z$ and $\hat{z}$. Recall that  $G_n(y)\rightarrow G(y)$, almost surely for all $y\geq 0$, then we  can write for $n$ large enough $G_n(y)=G(y)+ o(n^{-1/2}), \forall y\geq 0$. Thus, applying again the mean value theorem, we obtain for all large $n$,
\begin{eqnarray*}\label{p1}
 \frac{1}{n}\sum_{j=1}^{n}[\mathbb{I}(Y_{j,n}\leq \hat{z})-\mathbb{I}(Y_{j,n}\leq {z})]= G_n(\hat{z})-G_n(z)&=& G(\hat{z})-G(z)+o(n^{-1/2})\\
& =& g(z_1)[\hat{z}-z]+ o(n^{-1/2}),
\end{eqnarray*}
where $z_1$ is between $z$ and $\hat{z}$. Hence, the second term $II$ can be rewritten into
\begin{eqnarray*}
II&=&[\hat{z}-z]\frac{1}{n}\sum_{j=1}^{n} \frac{\partial}{\partial v} w[G_n(Y_{j,n}),G_n(z_0)]f(Y_{j,n},\hat{z})g(z_1)\mathbb{I}(Y_{j,n}\leq {z})\\
&+ &  [\hat{z}-z]\frac{1}{n}\sum_{j=1}^{n}\frac{\partial}{\partial z}f(Y_{j,n},z_0)w[G_n(Y_{j,n}),G_n(z)]\mathbb{I}(Y_{j,n}\leq {z})+ o(n^{-1/2}).
\end{eqnarray*}

 For the last term $III$, we also make use of  Taylor expansion.  For $Y_{j,n}$ in the vicinity of $z$, we have
\begin{eqnarray*}
 w[G_n(Y_{j,n}),G_n(\hat{z})]f(Y_{j,n},\hat{z})&=& w[G_n(z),G_n(\hat{z})]f(z,\hat{z})\\
 &+ &  \frac{\partial}{\partial u} w[G_n(z),G_n(\hat{z})]f(z,\hat{z})[G_n(Y_{j,n})-G_n(z)] \\
 &+& \frac{\partial}{\partial y}f(z,\hat{z})w[G_n(z),G_n(\hat{z})][Y_{j,n}-z]\\
 &+& o(|G_n(Y_{j,n})-G_n(z)|+|Y_{j,n}-z|).\\
\end{eqnarray*}
Thus $III$ can be transformed into
\begin{eqnarray*}
III&=&\frac{1}{n}\sum_{j=1}^{n}w[G_n(z),G_n(\hat{z})]f(z,\hat{z})[\mathbb{I}(Y_{j,n}\leq \hat{z})-\mathbb{I}(Y_{j,n}\leq {z})]\\
& +&\frac{1}{n}\sum_{j=1}^{n}\frac{\partial}{\partial u} w[G_n(z),G_n(\hat{z})]f(z,\hat{z})[G_n(\hat{z})-G_n(z)][\mathbb{I}(Y_{j,n}\leq \hat{z})-\mathbb{I}(Y_{j,n}\leq {z})] \\
& +& \frac{1}{n}\sum_{j=1}^{n}\frac{\partial}{\partial y}f(z,\hat{z})w[G_n(z),G_n(\hat{z})][Y_{j,n}-z][\mathbb{I}(Y_{j,n}\leq \hat{z})-\mathbb{I}(Y_{j,n}\leq {z})]\\
&+ & \frac{1}{n}\sum_{j=1}^{n}o(|G_n(Y_{j,n})-G_n(z)|+|Y_{j,n}-z|)[\mathbb{I}(Y_{j,n}\leq \hat{z})-\mathbb{I}(Y_{j,n}\leq {z})].
\end{eqnarray*}
Now, we are going show that the last three terms  in the right-hand side of the previous equality are asymptotically negligible.
For the second term we can write, in view of \eqref{p1}, that
 \begin{eqnarray*}
& \left|\frac{1}{n}\sum_{j=1}^{n}\frac{\partial}{\partial u} w[G_n(z),G_n(\hat{z})]f(Y_{j,n},\hat{z})[G_n(\hat{z})-G_n(z)][\mathbb{I}(Y_{j,n}\leq \hat{z})-\mathbb{I}(Y_{j,n}\leq {z})] \right|& \\
& \leq w[G_n(z),G_n(\hat{z})]f(z,\hat{z})|G_n(\hat{z})-G_n(z)|^2 &\\
& \leq w[G_n(z),G_n(\hat{z})]f(z,\hat{z})[g^2(z_1)|\hat{z}-z|^2+ o(n^{-1/2})].
 \end{eqnarray*}
 Thus, the second term tends to 0, because $\hat{z}\rightarrow z, \; n\rightarrow\infty$.
Analogously,  by using \eqref{p1} and assuming that the $Y_{j,n}$ are lying between $z$ and $\hat{z}$, we can major the third term by a quantity converging to 0.
\begin{eqnarray*}
& \left|\frac{1}{n}\sum_{j=1}^{n}\frac{\partial}{\partial y}f(z,\hat{z})w[G_n(Y_{j,n}),G_n(\hat{z})][Y_{j,n}-z][\mathbb{I}(Y_{j,n}\leq \hat{z})-\mathbb{I}(Y_{j,n}\leq {z})]\right| &\\
& \leq \frac{\partial}{\partial y}f(z,\hat{z})w[G_n(z),G_n(\hat{z})]|\hat{z}-z|[G_n(\hat{z})-G_n(z)] &\\
& \leq \frac{\partial}{\partial y}f(z,\hat{z})w[G_n(z),G_n(\hat{z})]g(z_1)|\hat{z}-z|^2 + o(n^{-1/2}).&\\  
\end{eqnarray*}
For the last and fourth term, one has
\begin{eqnarray*}
&\left|\frac{1}{n}\sum_{j=1}^{n}o(|G_n(Y_{j,n})-G_n(z)|+|Y_{j,n}-z|)[\mathbb{I}(Y_{j,n}\leq \hat{z})-\mathbb{I}(Y_{j,n}\leq {z})]\right|& \\
& \leq g(z_1)o(|\hat{z}-z|^2)+ o(n^{-1/2}).&
\end{eqnarray*}
Thus the fourth term also tends to 0, as $n\rightarrow\infty$. It follows from this that  $III$ is asymptotically equivalent to
\begin{equation}
III=w[G_n(z),G_n(\hat{z})]f(z,\hat{z})g(z_1)[\hat{z}-z)]+ o(n^{-1/2}).
\end{equation}
Finally, we obtain, for $n$ large enough, the following decomposition for  $\hat{J}_n(w,f)$ :
\begin{eqnarray*}
 \hat{J}_n(w,f) &= &\frac{1}{n}\sum_{j=1}^{n}w[G_n(Y_{j,n}),G_n({z})]f(Y_{j,n},{z}) \mathbb{I}(Y_{j,n}\leq {z})\\
&+ &[\hat{z}-z]\frac{1}{n}\sum_{j=1}^{n}\frac{\partial}{\partial v} w[G_n(Y_{j,n}),G_n(z_0)]f(Y_{j,n},\hat{z})g(z_0)\mathbb{I}(Y_{j,n}\leq {z})\\
&+ &[\hat{z}-z]\frac{1}{n}\sum_{j=1}^{n}\frac{\partial}{\partial z}f(Y_{j,n},z_0)w[G_n(Y_{j,n}),G_n(z)]\mathbb{I}(Y_{j,n}\leq {z})\\
 &+& w[G_n(z),G_n(\hat{z})]f(z,\hat{z})g(z_1)[\hat{z}-z)]+ o(n^{-1/2}).
\end{eqnarray*}
Combining the continuity of the partial derivatives of $w$ and $f$ on the compact interval $[0,z]$ and the fact that $z_0$ converges to $z$, we can approximate the second and the third summation terms in the second member of the above equality  respectively by the integrals $$\int_{0}^{z}\frac{\partial}{\partial v} w[G(y),G(z)]f(y,z)g(z)dG(y)$$ and $$\int_{0}^{z}\frac{\partial}{\partial z}f(y,z) w[G(y),G(z)]dG(y).$$  
Besides, since $\hat{z}$ converges almost surely to $z$ and $f$ continuous,  $f(z,\hat{z})$ converges almost surely to $f(z,z)$, which is equal to 0, in virtue of the normative focalisation axiom on poverty indices.  Thus, the fourth  term converges to 0, and $\hat{J}_n(w,f)$ becomes
\begin{equation}\label{r1}
\hat{J}_n(w,f)={J}_n(w,f)+a(\hat{z}-z)+ o(n^{-1/2}),
\end{equation} 
where 
$$a=\int_{0}^{z}\left(\frac{\partial}{\partial v} w[G(y),G(z)]f(y,z)g(z)+ \frac{\partial}{\partial z}f(y,z) w[G(y),G(z)]\right) dG(y).$$
From this, we can conclude that $\hat{J}_n(w,f)$ converges almost surely to ${J}(w,f)$, as $n\rightarrow\infty$.$\square$\\
 
 Now, we state our main result  which is the weak convergence of the normalized and centered process $\{\sqrt{n}[\hat{J}_{n}(w,f)-J(w,f)]: w\in \mathcal{W}, f\in\mathcal{F}\}$ in $l^{\infty }(\mathcal{W}\times
\mathcal{F})$, the set of all real-valued and bounded functions defined on $\mathcal{W}\times \mathcal{F}.$
\begin{theorem}\label{t1}
Let $G(y)$ be a continuous distribution function with probability density  $g(y)$. If assumptions (A.1-2)  hold, then the process
$\{\sqrt{n}[\hat{J}_{n}(w,f)-J(w,f)]: w\in \mathcal{W}, f\in\mathcal{F}\}$ converges weakly in $l^{\infty }(\mathcal{W}\times\mathcal{F})$ to a zero-mean Gaussian process with covariance function defined, for any $(w,f),(\widetilde w,\widetilde f)\in\mathcal{W\times F}$, as
\begin{eqnarray*}\label{st}
\Gamma \left[(w,f); (\widetilde w,\widetilde f)\right] &=& \Sigma \left[(w,f); (\widetilde w,\widetilde f)\right] +a\left( \int_{0}^{\infty} h(y)\zeta(y)dG(y)-\mathbb{E}[\zeta(Y)]J(w,f)\right)  \\ \nonumber
& +  & a\left( \int_{0}^{\infty} \widetilde{h}(y)\zeta(y)dG(y)-\mathbb{E}[\zeta(Y)]J(w,f)\right)+ a^2{\rm Var}[\zeta(Y)],
\end{eqnarray*}
with 
$$ h(y)=w[G(y),G(z)]f(y,z)\mathbb{I}(y\leq z) \quad; \quad\widetilde{h}(y)=\widetilde w[G(y),G(z)]\widetilde f(y,z)\mathbb{I}(y\leq z)$$
and
\begin{eqnarray*}
\Sigma \left[(w,f); (\widetilde w,\widetilde f)\right] &=&
\int^{z}_{0}w[G(y),G(z)]f(y,z)\widetilde w[G(y),G(z)]\widetilde f(y,z)dG(y)\\
 &- &\int^{z}_{0}w[G(y),G(z)]f(y,z)dG(y)\int^{z}_{0}\widetilde w[G(y),G(z)] \widetilde f(y,z)dG(y)\\
 &+&\int_0^{z}\int_0^{z} a_1(x,y)[G(x)\wedge G(y) - G(x)G(y)]dG(x)dG(y), \nonumber  \\
&+& [1-G(z)]\int_0^{z}\int_0^{z} a_2(x,y)G(y)dG(x)dG(y) \nonumber  \\
&+&[1- G(z)] \int_0^{z}\int_0^{z} a_3(x,y)G(x)dG(x)dG(y)  \nonumber  \\
&+&  G(z)[1-G(z)] \int_0^{z}\int_0^{z} a_4(x,y)dG(x)dG(y), \nonumber 
\end{eqnarray*} 
where
\begin{eqnarray*}
a_1(x,y) & =& \frac{\partial w}{\partial u}[G(x),G(z)]f(x,z)\frac{\partial \widetilde w}{\partial u}[G(y),G(z)] \widetilde f(y,z),\\
a_2(x,y) & =& \frac{\partial w}{\partial
u}[G(x),G(z)]f(x,z)\frac{\partial \widetilde w}{\partial v}[G(y),G(z)] \widetilde f(y,z),\\
a_3(x,y)& =&\frac{\partial w}{\partial v}[G(x),G(z)]f(x,z)\frac{\partial \widetilde w}{\partial u}[G(y),G(z)] \widetilde f(y,z),\\
a_4(x,y)& =& \frac{\partial w}{\partial v}[G(x),G(z)]f(x,z)\frac{\partial \widetilde w}{\partial v}[G(y),G(z)] \widetilde f(y,z).\\
\end{eqnarray*}
\end{theorem}
\textbf{Remark 1.} For any given functions $w$ and $f$, Theorem \ref{t1} gives the variance of the estimator $\hat{J}_n(w,f)$ which is equal to
\begin{eqnarray}\label{stt}
\Gamma \left[(w,f); (w, f)\right] &=& \Sigma \left[(w,f); ( w,f)\right] +2a\left(\int_{0}^{\infty} h(y)\zeta(y)dG(y) -\mathbb{E}[\zeta(Y)]J(w,f)\right)  \nonumber\\
 &&+ a^2{\rm Var}[\zeta(Y)]. 
\end{eqnarray}
This means that the variance of the poverty estimate $\hat{J}_n(w,f)$ is increased by a term $$\Delta=2a\left(\int_{0}^{\infty} h(y)\zeta(y)dG(y)-\mathbb{E}[\zeta(Y)]J(w,f)\right)+ a^2{\rm Var}[\zeta(Y)]$$ whenever the poverty line $z$ is estimated from the sample.\\

\textbf{Remark 2.} If the function $w\equiv 1$, which corresponds to the additively decomposable measures, one can observe that all the integrals  with $a_i$-term $i=1,2,3,4$  vanish.Then the remainding terms in \eqref{stt} are exactly the expressions in Equations (21) and (22) in \cite{zh2} for suitable functions $\zeta$. The quantity corresponding to sum of all  integrals  with $a_i$-term is due to the weight function $w$, 
when it is considered, as for example, in the case of Sen poverty index.\\

\textbf{Proof.} It relies on the following decomposition which is obtained from \eqref{r1}
\begin{equation}\label{r2}
\sqrt{n}[\hat{J}_n(w,f)-J(w,f)]=\sqrt{n}[{J}_n(w,f)-J(w,f)]+ \frac{a}{\sqrt{n}}\sum_{j=1}^{n}\zeta(Y_j) +o_{\mathbb{P}}(1).
\end{equation}
Observe that the  term $\frac{a}{\sqrt{n}}\sum_{j=1}^{n}\zeta(Y_j)$ in the right-hand side of \eqref{r2} is a sum of independent and identically random variables with mean $\mathbb{E}[\zeta(Y)]$ and finite variance ${\rm Var}[\zeta(Y)]=\int \zeta^2(y)dG(y)$. Then, by applying the central limit theorem, it converges in law to a Gaussian random variable, with  variance $a^2 {\rm Var}[\zeta(Y)]$.\\
\indent Next, by using the modern theory of empirical processes indexed by functions, we prove in Appendix (see, also \cite{ls}) that the centered and normalized process $\{\sqrt{n}[J_{n}(w,f)-J(w,f)]: w\in \mathcal{W}, f\in\mathcal{F}\}$ converges weakly in $l^{\infty }(\mathcal{W}\times\mathcal{F})$ to a tight Gaussian process with zero-mean and covariance function given by the kernel $\Sigma(\cdot,\cdot)$. This entails that the process $\{\sqrt{n}[J_{n}(w,f)-J(w,f)]: w\in \mathcal{W}, f\in\mathcal{F}\}$ is asymptotically tight. Since the second term $\frac{a}{\sqrt{n}}\sum_{j=1}^{n}\zeta(Y_j)$ does not depend on the indexing parameter $(w,f)$, we can infer that the sum process $\{\sqrt{n}[{J}_{n}(w,f)-J(w,f)]+\frac{a}{\sqrt{n}}\sum_{j=1}^{n}\zeta(Y_j): w\in \mathcal{W}, f\in\mathcal{F}\}$ is asymptotically tight.
 Moreover,  the $o_{\mathbb{P}}(1)$-term converges uniformly to 0 in $(w,f)$, as it does  not depend on $(w,f)$.  Thus, since the finite margins of this process are Gaussian (by applying the multivariate central limit theorem), we can conclude that $\{\sqrt{n}[\hat{J}_{n}(w,f)-J(w,f)]: w\in \mathcal{W}, f\in\mathcal{F}\}$ in distribution to a limit Gaussian process. By independence of the $Y_{j}$'s, the cross covariance of the two terms in the right hand side of \eqref{r2} is given for all $(w,f)$ by
\begin{eqnarray*}
\sigma_{w,f,\zeta} &=&\frac{a}{n} \sum_{i=1}^{n}\sum_{j=1}^{n}{\rm cov}(w[G_n(Y_{i,n}),G_n({z})]f(Y_{i,n},{z}), \zeta(Y_j))\\
&=& a.{\rm cov}(w[G_n(Y_{i,n}),G_n({z})]f(Y_{i,n},{z}), \zeta(Y_j))\\
&=& a\left(\int_{0}^{z}w[G_n(y),G_n(z)]f(y,{z}) \zeta(y)dG(y) \right.\\
 &  & \left. -\int_{0}^{z}w[G_n(y),G_n(z)]f(y,{z})dG(y)\int_{0}^{\infty}\zeta(y)dG(y)\right)\\
&\longrightarrow & a\left(\int_{0}^{z}w[G(y),G(z)]f(y,{z}) \zeta(y)dG(y)-\mathbb{E}[\zeta(Y)]J(w,f)\right),\; n \rightarrow\infty.
\end{eqnarray*} 

 \section{ Testing procedures }
 Inference procedures for  testing poverty  usually allow one to say that there is less or more poverty in a given population than in another, but  do not permit to answer questions of type : \textit{How much poverty has been decreased or increased ?} Therefore, it is not possible to use  these procedures in order to quantify the poverty variation (or change) between two populations . In this section, we propose a testing procedure which allows us to evaluate the poverty change between two populations, by checking  for whether there exists a proportionality relation between their associated poverty indices. That is, we aim to test the following hypotheses :
   $$H_0: J_F=\alpha J_G,\qquad \text{versus}\qquad  H_1: J_F\neq \alpha J_G,$$
where $\alpha$ is a positive real number, and  $J_F$,  $J_G$ are aggregated scalar poverty  indices computed from distributions $F$ and $G$. Note that $J_F$ and $J_G$ must be defined with the same specific functions $w$ and $f$ indicating the type of poverty measure being considered.
The acception of the null hypothesis $H_0$ leads to the estimation of the relative poverty variation between the two distributions $F$ and $G$. For example, if $\alpha=1/2$, we can say that poverty has decreased by an half, if the reference distribution is $G$.  These kind of conclusions are important for policy makers, as they enable to show the effect of poverty reducing strategies. Indeed, our approach  may be applied to check for the poverty Millennium Development Goals (MDG) which consisted of halving poverty in the world by 2015.\\

\indent Consider now two independent populations with cumulative distribution functions $F$ and $G$, and relative poverty lines $z_F$ and $z_G$, respectively. Assume that two independent and identically distributed samples of sizes $n_F$ and $n_G$ are respectively drawn from thereof. Denote by $\widehat{J}_F=\widehat{J}_F(w,f)$ and $\widehat{J}_G=\widehat{J}_G(w,f)$ the estimators of $J_F$ and $J_G$, respectively. By
{Theorem \ref{t1}}, $\widehat{J}_F$ and $\widehat{J}_G$ are
asymptotically normally distributed, with  variances $\sigma^2_F$ and $\sigma^2_G$ that can be readily determined from \eqref{st} by computing $\Gamma[(w,f);(w,f)]$ with the right distribution $F$ or $G$.\\

Let $\Delta \widehat{J}=\widehat{J}_F- \alpha\widehat{J}_G$. Then, under $H_0$, $\Delta \widehat{J}$ is
asymptotically normally distributed with zero mean and variance $\sigma^2$ which,
by independence of the two samples, is equal to
$$\sigma^2=\frac{1}{n_F}\sigma^2_F +\frac{\alpha^2}{n_G}\sigma^2_G.$$ 
A consistent estimator for $\sigma^2$, may be defined as
$$
\widehat{\sigma}^2=\frac{1}{n_F}\widehat{\sigma}^2_F+\frac{\alpha^2}{n_G}\widehat{\sigma}^2_G,
$$
where $\widehat{\sigma}^2_F$ and $\widehat{\sigma}^2_G$ are consistent estimators for ${\sigma}^2_F$ and ${\sigma}^2_G$, respectively. Thus, for checking the null hypothesis $H_0$, we may use the following test statistic :
$$ \widehat{T}=\frac{\Delta \widehat{J}}{\widehat{\sigma}},$$
which, by Slutsky's Theorem, converges in law to the standard normal distribution, under $H_0$.\\

\indent The previous test is distribution-free and may be extended to a vector of several particular poverty indices. To perform a joint test using simultaneously several poverty indices, we consider two finite $d$-dimensional vectors ($d$ is a positive integer) of poverty indices denoted by $I_F$ and $I_G$ and associated with distributions $F$ and $G$, respectively. The hypotheses we wish to test are  of the form :
$$H_0: I_F= M I_G,\qquad \text{versus}\qquad  H_1: I_F\neq M I_G,$$
where $M={\rm diag}(\alpha_1,\cdots,\alpha_d)$ is a diagonal matrix of positive coefficients $\alpha_i$. When the matrix $M$ coincides with the identity matrix $\textbf{I}_d$, hypothesis $H_0$ corresponds to the equivalence of the two distributions in terms of poverty. The test can performed by making use of the Wald test statistic which is defined as
$$ \widehat{W}= (\widehat{I}_F-M \widehat{I}_G)'\widehat{\Gamma}^{-1}_{FG}(\widehat{I}_F-M \widehat{I}_G), $$ 
where $x'$ designs the transpose of a vector $x\in\mathbb{R}^{d}$,$\; \widehat{I}_F, \widehat{I}_G$ are consistent estimators of ${I}_F$ and ${I}_G$, respectively.  $\widehat{\Gamma}_{FG}$ is a consistent estimator of the covariance matrix of the vector $I_F-M I_G$.  It is clear that, under $H_0$,  the statistic $\widehat{W}$  converges weakly  to a chi-square distribution with $d$ degrees of freedom, $\chi^{2}(d)$. Thus, at level of significance $\alpha$, the critical region is of the form $\{\widehat{W} > c\},$ where $c$ is the $(1-\alpha)$-quantile of $\chi^{2}(d)$.\\
 
 \indent To estimate the covariance matrix $\Gamma$ in Theorem \ref{t1}, denote $\widehat{\Gamma}_{k,l}$ the estimators of the entries ${\Gamma}_{kl}:=\Gamma[(w_l,f_l),(w_k,f_k)],\, k,l=1,\cdots,d$. Take $z$ equal to the quantile of order $q/n$ of a the considered distribution $F$ or $G$, represented by the sample  $Y_1,\cdots,Y_n$, with $q$ a positive  integer less than $n$. Then, the number of poor in the sample is equal to $q$, and a consistent estimators of the elements $\Sigma_{k,l}:=\Sigma\left[(w_{k},f_{k}); (w_{l},f_{l})\right]$,  $1\leq k,l\leq d$, may be defined as
 \begin{eqnarray} \label{eq:om}
\widehat{\Sigma}_{k,l} &= &
\frac{1}{n}\sum_{j=1}^{q}w_k\left(\frac{j}{n},\frac{q}{n}\right)f_k(Y_{j,n},z)w_l\left(\frac{j}{n},\frac{q}{n}\right)f_l(Y_{j,n},z) \nonumber \\
 &- & \frac{1}{n^2}\sum_{i=1}^{q}\sum_{j=1}^{q}w_k\left(\frac{i}{n},\frac{q}{n}\right)f_k(Y_{i,n},z)w_l\left(\frac{j}{n},\frac{q}{n}\right)f_l(Y_{j,n},z) \nonumber \\
& + & \frac{1}{n^2}\sum_{i=1}^{q}\sum_{j=1}^{q}a_1(Y_{i,n},Y_{j,n})\left(\frac{i}{n}\wedge\frac{j}{n}-\frac{i}{n}\frac{j}{n}\right)\nonumber \\
& +& \left(1-\frac{q}{n}\right) \frac{1}{n^2}\sum_{i=1}^{q}\sum_{j=1}^{q}a_2(Y_{i,n},Y_{j,n})\frac{j}{n}  \nonumber \\
& + & \left(1-\frac{q}{n}\right)\frac{1}{n^2}\sum_{i=1}^{q}\sum_{j=1}^{q}a_3(Y_{i,n},Y_{j,n})\frac{i}{n} \nonumber \\ 
& + &\frac{q}{n}\left(1-\frac{q}{n}\right)\frac{1}{n^2}\sum_{i=1}^{q}\sum_{j=1}^{q}a_4(Y_{i,n},Y_{j,n}),
\end{eqnarray}

where for all $i$, $w_i$ and $f_i$ represent respectively the weighting and deprivation functions of a particular poverty index, $a_r, r=1,2,3,4$ are real-valued functions given in Theorem \ref{t1},  and $Y_{1,n}\leq\cdots\leq Y_{n,n}$
are the order statistics associated with the sample $Y_{1},\cdots,Y_{n}$. It follows from this, that consistent estimators for the entries $\Gamma_{k,l},\, 1\leq k,l\leq d$, are 
\begin{eqnarray}\label{vaR}
\widehat{\Gamma}_{k,l}&=&\widehat{\Sigma}_{k,l}+a\left(\frac{1}{n}\sum_{j=1}^{n}h_k(Y_{j,n})\zeta(Y_{j,n}) -\frac{1}{n^2}\sum_{j=1}^{n}h_k(Y_{j,n})\sum_{j=1}^{n}\zeta(Y_{j,n})\right) \nonumber \\
&+&a\left( \frac{1}{n}\sum_{j=1}^{n}h_l(Y_{j,n})\zeta(Y_{j,n})-\frac{1}{n^2}\sum_{j=1}^{n}h_l(Y_{j,n})\sum_{j=1}^{n}\zeta(Y_{j,n})\right) \nonumber \\
&+& \frac{a^2}{n}\sum_{j=1}^{n}\left(\zeta(Y_{j,n})-\frac{1}{n}\sum_{j=1}^{n}\zeta(Y_{j,n})\right)^2.
\end{eqnarray}
which lead to a consistent and non-parametric estimator for the covariance matrix $\Gamma$.

\section{Simulation study}
Here, we make some experiments for showing the asymptotic normality of our estimator $\hat{J}_n$ in relatively small samples of sizes $n=50,100,150.$
We essentially deal with two simple cases ; that is the case where the relative poverty line $z$ is taken equal to the mean of the distribution and the case where $z$ is set to the median of the distribution. The simulation procedure is the following :
\begin{itemize}
\item Generate data from a known distribution with positives values ;
\item Calculate the estimator $\hat{J}_n(w,f)$ and the theoretical indice $J(w,f)$ for specific functions $w,f$ ;
\item Compute the variance, say $\sigma^2_n(w,f)$, by using \eqref{vaR} ;
\item Determine the statistic test $T_n=\left|\frac{\hat{J}_n(w,f)- J(w,f)}{\sigma_n(w,f) }\right|$ ;
\item compute the p-value $p=2*(1-\phi(T_n))$, where $\phi$ is the standard Gaussian distribution;
\item Repeat all these steps $B$ times.
\end{itemize}
We generate  data from two distributions : Exponential(1/2) and Lognormal(0,1). In case of the  poverty line being equal to the mean of the distribution, the function $\zeta(y)=y$. While the median case corresponds to $\zeta(y)=\frac{1}{g(G^{-1}(1/2))}\mathbb{I}(y\leq G^{-1}(1/2))$. The tables below present the p-values of the normality test for different indices : FGT(1), FGT(2) and Sen, by considering a number of replications $B=1000$.  

\begin{table}[htbp]
$$\begin{array}{|c|c|c|c||c|c|c|}
\hline
       &\multicolumn{3}{c||}{\text{Exponential}(1/2)} &  \multicolumn{3}{c|}{\text{Lognormal}(0,1)} \\
\hline
\text{Size} & \text{FGT(1)} &  \text{FGT(2)} & \text{Sen} &  \text{FGT(1)} &  \text{FGT(2)} & \text{Sen}\\
 \hline
n=50& 0.87   &0.86    &0.92   &0.89   &0.93    &0.72\\
 \hline
n=100&  0.93  &0.92    & 0.94   &0.91   &0.96    &0.81\\
 \hline
n=150& 0.95   & 0.93 & 0.94  &0.99   & 0.99   &0.89\\
\hline
\end{array}
$$
\caption{$p$-values of test in case the poverty line $z$ is equal to the mean of the distribution.}
\label{tab1}
\end{table}

\begin{table}[htbp]
$$\begin{array}{|c|c|c|c||c|c|c|}
\hline
       &\multicolumn{3}{c||}{\text{Exponential}(1/2)} &  \multicolumn{3}{c|}{\text{Lognormal}(0,1)} \\ 
\hline
\text{Size} & \text{FGT(1)} &  \text{FGT(2)} & \text{Sen} &  \text{FGT(1)} &  \text{FGT(2)} & \text{Sen}\\
 \hline
n=50& 0.24 &0.33    &0.37    &0.03   &0.21   &0.06\\
 \hline
n=100&  0.62  &0.74   & 0.80   &0.08   &0.40    &0.20\\
 \hline
n=150& 0.86  & 0.88   & 0.95  &0.17  & 0.62   &0.36\\
\hline
\end{array}
$$
\caption{$p$-values of test in case the poverty line $z$  is equal to the median of the distribution.}
\label{tab2}
\end{table}

Tables \ref{tab1} and \ref{tab2} allow us to accept the asymptotic normality of  our estimator for sample sizes greater than or equal to $n=100$ at a nominal level of 5\%.

\section{Application to genuine data}
Here, we use real data to estimate the quantity $\Delta$, representing the added term to the variance of the poverty estimate when we deal with  a relative poverty line $z$.
We employ Senegalese households expenditure data, which consist of a sample of size $n=3163$ provided  by a national survey entitled \textit{ESAM 2}, that was conducted in 2001 by the National Statistic Agence (ANSD). We consider two relative poverty lines : the mean and the median of the observed distribution. Denote the data by 
$y_1\leq y_2\leq \cdots \leq y_n$. Then, an estimation of the quantity $\Delta$ is given by
\begin{eqnarray*} 
\hat{\Delta}&=& 2\hat{a}\left[\frac{1}{n}\sum_{i=1}^{q} w\left(\frac{i}{n},\frac{q}{n}\right)f(y_i,z)\zeta(y_i) - \overline{\zeta(Y)}. \frac{1}{n}\sum_{i=1}^{q} w\left(\frac{i}{n},\frac{q}{n}\right)f(y_i,z)\right]\\
&& + \hat{a}^2\left[\frac{1}{n}\sum_{i=1}^{n}\left(\zeta(y_i)-\overline{\zeta(Y)}\right)^2\right],
\end{eqnarray*}
where 
$$ \hat{a}= \frac{1}{n}\sum_{i=1}^{q}\left[ \frac{\partial}{\partial v} w\left(\frac{i}{n},\frac{q}{n}\right)f(y_i,z) +\frac{\partial}{\partial z} f(y_i,z)w\left(\frac{i}{n},\frac{q}{n}\right)\right],$$
$\overline{\zeta(Y)}=\frac{1}{n}\sum_{i=1}^{n}\zeta(y_i)$ and $q$ is the number of poor in the sample and satisfies : $G_n(z)=q/n$.\\
The following table \ref{tab3} gives estimations for the quantity $\Delta$ and the variance of the poverty estimate when the poverty line is fixed. The results concern the Sen index and the FGT indices of parameter $\beta=1,2.$\\

\begin{table}[htbp]
$$
\begin{array}{|c|c|c|c|}
\hline
    & \text{FGT(1)} &  \text{FGT(2)} &  \text{Sen}  \\
 \hline
\multirow{2}{*}{z=\text{mean}}& \hat{\Delta}=0.004 & \hat{\Delta}=0.01  & \hat{\Delta}=-0.05       \\
                                               & \hat{var}=0.07 &  \hat{var}=0.03   & \hat{var}=0.23 \\
\hline
\multirow{2}{*}{z=\text{median}}& \hat{\Delta}=1.44 & \hat{\Delta}=0.38  &\hat{\Delta}=1.63     \\
                                                   & \hat{var}=0.05  &     \hat{var}=0.01  &  \hat{var}=0.15 \\
\hline
\end{array}
$$
\caption{Estimation of $\Delta$ and  variances of poverty estimates when the poverty line is fixed.}
\label{tab3}
\end{table}

We observe that when the poverty line $z$ is set to the median of the distribution, the sampling error due the estimation of $z$ i increases the variance of the poverty estimate  for all the considered indices. In contrast, when the poverty line is set to the mean of the distribution, the variance of the FTG indices increases while the variance of the Sen measure decreases. This may be due to the fact that, the Sen measure affects a weight which depends on the poverty line $z$ and the individual's ranks.

\section*{Appendix}
\subsection{Proof of Theorem \ref{t1} }
First, recall the definition of the classes of functions $\mathcal{W}$ and $\mathcal{F}$
$$\mathcal{W}=\{w:[0,1] \times [0,1] \rightarrow \mathbb{R}_{+}, \; w
\;\;\text{  continuous,}\;\;\text{and}\;\;u\mapsto w(u,\cdot)
\;\;\text{is non-increasing}\}$$ 
$$\mathcal{F}=\{f:\mathbb{R}_{+}\times\mathbb{R}_{+}\rightarrow \mathbb{R}_{+},\;f
\;\;\text{ continuous,}\;\;\text{and}\;\;y\mapsto f(y,\cdot)
\;\;\text{is non-increasing}\}$$
Next, introduce the  class of functions
$$\mathcal{K}=\{k:\mathbb{R} \rightarrow [0,1] \;  \text{increasing} \}.$$
For  $z>0$ fixed, $ w\in\mathcal{W},f\in\mathcal{F}, k \in\mathcal{K}$  define the real-valued function
$$ h_{w,f,k}(y) = w[k(y),k(z)]f(y,z) \mathbb{I}(y <z),\quad\text{for all}\; y\in\mathbb{R}_{+}$$
and let $\mathcal{H}_z $ be the class of functions defined as
$$
\mathcal{H}_z  =  \{y\mapsto h_{w,f,k}(y) : w\in\mathcal{W},f\in\mathcal{F}, k \in\mathcal{K}\}.
$$
According to the sketch given at the end of the statement of the Theorem \ref{t1}, we split the proof  into four parts. In the first, we establish the Donsker property for the class $\mathcal{H}_z$, and derive from this, that the empirical process $\{\mathbb{G}_n(h_{w,f,G}) : w\in\mathcal{W},f\in\mathcal{F}\}$ converges weakly 
to a limit Gaussian process $\mathbb{G}(h_{w,f,G})$.
In the second, we show that 
\begin{equation}\label{84'}
\sup_{(w,f) \in \mathcal{W}\times\mathcal{F}}|\mathbb{G}_{n}(h_{w,f,G_n}-h_{w,f,G})|\longrightarrow_{p} 0,\quad n\rightarrow\infty,
\end{equation}
where "$\longrightarrow_{p}$" denotes the convergence in probability.
In the third part, we prove the weak convergence of the process $\mathbb{W}_n(w,f)$ to a zero-mean Gaussian process $\mathbb{W}(w,f)$ in $l^{\infty}(\mathcal{W}\times\mathcal{F})$. Finally in the last part, we prove that the joint process $(\mathbb{G}_n,\mathbb{W}_n)$ converges weakly to $(\mathbb{G},\mathbb{W})$ which is a zero-mean Gaussian process. 

 \subsection{Part I}
Recall that $P$ is the common probability law of the $Y_j's$ and $G$ stands for its cumulative distribution function. 
We have to prove that the class of functions $\mathcal{H}_z$ is $P$-Donsker. This will be done if we prove that  the bracketing integral
 $$
 J_{[]}(\infty,\mathcal{H}_z,L_2(P))=\int_0^{\infty}\sqrt{\log
N_{[]}(\epsilon,\mathcal{H}_z, L_{2}(P))}d\epsilon
 $$
 is finite, where $N_{[]}(\cdot)$ denotes the bracketing number. Before proving this, observe that the elements of 
$\mathcal{H}_z$ are continuous and increasing functions, bounded on $\mathbb R_+$ by $w[k(0), k(z)]f(0,z)$, for every $(w,f,k) \in \mathcal {W \times F \times K}$. By assumption (A), the classes of functions $\mathcal{W}$ and $\mathcal{F}$ are pointwise measurable.  
Further, Lemma 2.2 of \cite{vdg} entails that the $\delta$-entropy, relatively to the supremum norm,  of the class of increasing functions $\mathcal{K}$ is finite for any $\delta>0$.
 That is, the class $\mathcal K$ is totally bounded relatively to the supremum norm, and hence is pointwise measurable. 
This enables us to take the supremum over the set $\mathcal{W} \times \mathcal{F}\times \mathcal{K}$ as equal to the supremum over a countable subset $\mathcal{G}_0\subset\mathcal{W} \times \mathcal{F}\times \mathcal{K}$. Since for  $z>0$ fixed, the quantity $w[k(0), k(z)]f(0,z)$ is finite for any $(w,f,k) \in \mathcal {W \times F \times K}$, we may define the constant function 
$$
H(y)=\sup_{(w,f,k) \in \mathcal{W} \times \mathcal{F}\times \mathcal{K}}w[k(0), k(z)]f(0,z),\;\forall y\in\mathbb R_+,
$$ 
as an envelope function for the class $\mathcal{H}_z$. Then $\mathcal{H}_z$ is uniformly bounded by $H(y)$, and we may assume without loss of generality that 
$H(y)\equiv 1$. Thus, $\mathcal{H}_z$ is a subset of the class of monotone functions defined on $\mathbb R$ with values in $[0,1]$. It follows from Theorem 2.7.5, page 159 of \cite{vdv} that for all $\epsilon >0$, 
 \begin{equation}\label{sw00}
\log N_{[]}(\epsilon,\mathcal H_z, L_{2}(P))< C\epsilon^{-1},
\end{equation}
where $C$ is a positive constant.\\
From the fact that the elements of $\mathcal{H}_z$ take their values in $[0,1]$, for $\epsilon > 1 $ the number of $\epsilon$-brackets needed to cover $\mathcal{H}_z$ is just 1. Then $J_{[]}(\infty,\mathcal{H}_z,L_2(P))$ would be finite if $$ \int_0^{1}\sqrt{\log
N_{[]}(\epsilon,\mathcal{H}_z, L_{2}(P))}d\epsilon < \infty.$$
 Now, integrating both sides of (\ref{sw00}), one obtains  
$$
\int_0^{1}\sqrt{\log
N_{[]}(\epsilon,\mathcal{H}_z, L_{2}(P))}d\epsilon< \sqrt{C}\int_0^1 \epsilon^{-1/2}d\epsilon=2\sqrt{C}<\infty.
$$
 That is, $J_{[]}(\infty,\mathcal{H}_z,L_2(P))$ is finite and the class $\mathcal{H}_z$ is $P$-Donsker. 
In particular for $k=G$ (the distribution function  associated with the probability law $P$), the class  $\mathcal{H}_{z}$ restricts to
$$
\mathcal{H}_{z,G}=\{h_{w,f,G}:w\in\mathcal{W},f\in\mathcal{F}\},
$$
which may be identified to $\mathcal{W}\times\mathcal{F}$. Since $\mathcal{H}_{z,G}\subset\mathcal{H}_z$ is $P$-Donsker, so is the class $\mathcal{W}\times\mathcal{F}$. Then it follows that the empirical process $\{\mathbb{G}_n(h_{w,f,G}) : w\in\mathcal{W},f\in\mathcal{F}\}$
converges weakly in $l^{\infty}(\mathcal{H}_{z,G})$ to a tight limit process $\mathbb{G}$ , which is a zero- mean Gaussian process with covariance function defined, for all $(w,f)$ and $(\widetilde{w},\widetilde{f})$, by
\begin{eqnarray*}
 {\rm cov}(\mathbb{G}(h_{w,f,G}),\mathbb{G}(h_{\widetilde{w},\widetilde{f},G}))
&= & Ph_{w,f,G}h_{\widetilde{w},\widetilde{f},G}-Ph_{w,f,G}Ph_{\widetilde{w},\widetilde{f},G}\\
& = & \int^{z}_{0}w[G(y),G(z)]f(y,z)\widetilde{w}[G(y),G(z)]\widetilde{f}(y,z)dG(y) \\
 &-&\int^{z}_{0}w[G(y),G(z)]f(y,z)dG(y) \int^{z}_{0}\widetilde{w}[G(y),G(z)]\widetilde{f}(y,z)dG(y).
\end{eqnarray*}
 
 \subsection{Part II}
For establishing (\ref{84'}), we first remark that 
%
for any $(w,f) \in \mathcal W \times \mathcal F$, the functions $h_{w,f,G}$  and $h_{w,f,G_n}$ are elements of $\mathcal H_z$, which is shown to be $P$-Donsker according to the preview part. Since $h_{w,f,G}$  and $h_{w,f,G_n}$ are bounded, they are in $L_2(P)=L_2(G)$.  Now, one has 
\begin{eqnarray*}
\int_0^{\infty} \left[h_{w,f,G_n}(y)- h_{w,f,G}(y) \right]^2 dG(y) & \le & \sup_{y \le z} \left[h_{w,f,G_n}(y)- h_{w,f,G}(y) \right]^2\\
&\leq &\sup_{y \le z} |w(G_n(y),G_n(z))-w(G(y),G(z))|^2f^2(y,z)\\
&\le & f^2(0,z)\sup_{y \le z} |w(G_n(y),G_n(z))-w(G(y),G(z))|^2,
\end{eqnarray*}
which tends almost surely to 0, as $n\rightarrow\infty$, by  continuity of the function $w$ and the fact that
the empirical distribution function $G_n(y)$ converges almost surely to $G(y)$ for all $y\in\mathbb{R}$. Thus, as $n$ tends to infinity, 
$\int_0^{\infty} \left[h_{w,f,G_n}(y)- h_{w,f,G}(y) \right]^2 dG(y)$ 
 converges almost surely and hence in probabilty to zero. It follows from Lemma 19.24 of \cite{vdv1} that $\mathbb{G}_n(h_{w,f,G_n}- h_{w,f,G}){\longrightarrow}_p 0$, $n \rightarrow \infty$ which, by the continuous mapping theorem, implies that 
$$ \sup_{(w,f) \in \mathcal W \times \mathcal F}| \mathbb{G}_n(h_{w,f,G_n}- h_{w,f,G})|{\longrightarrow}_p 0, \ n \rightarrow \infty.$$ 
This establishes the second part of our proof.

\subsection{Part III}
 For any given functions $(w,f)\in\mathcal{W\times F}$, we define on the class $\mathcal K=\{k:\mathbb{R} \rightarrow [0,1] ,\;  \text{increasing} \}$ the following operator 
 $$\phi_{w,f} : k \mapsto \phi_{w,f}(k)=\int_0^z w[k(y),k(z)]f(y,z)dP(y)=P h_{w,f,k},$$

Recall that 
${\partial \over \partial u} \xi(a,b)$ and ${\partial \over \partial v}\xi(a,b)$ are the partial derivatives of a differentiable function $\xi(u,v)$ with respect to its first and second arguments, taken at $(u,v)=(a,b)$. 
Let $$\mathcal{K}'=\{k\in\mathcal{K},\;k \;\text{ continuous} \}.$$
For all $k \in \mathcal K$, and $s_t \in \mathcal K$ such that $k+ts_t \in \mathcal K$ and $s_t \rightarrow s \in \mathcal K',$ as $ t \rightarrow 0$, one has by a first-order Taylor expansion of $w$, for some functions $\zeta$ and $\pi$ defined on $\mathbb R$,  with values in $ (0,1)$ :
\begin{eqnarray*}
 { \phi_{w,f}(k+ts_t) -  \phi_{w,f}(k) \over t}
&=& 
\int_0^z  s_t(y) {\partial \over \partial u}w \left[k(y)+t\pi(t)s_t(y),k(z)+t\zeta(t)s_t(z)\right] f(y,z)dG(y)\\
&&+\int_0^z   s_t(z) {\partial \over \partial v}w\left[k(y)+t\pi(t)s_t(y),k(z)+t\zeta(t)s_t(z)\right] f(y,z)dG(y) \\
&=:& I_t+II_t. 
\end{eqnarray*}
Now, we have to show that as $t \rightarrow 0$, 
\begin{eqnarray*}
I_t & \longrightarrow & I=\int_0^z  s(y) {\partial \over \partial u}w \left[k(y),k(z)\right] f(y,z)dG(y) \\ 
II_t & \longrightarrow & II=\int_0^z   s(z) {\partial \over \partial v}w\left[k(y),k(z)\right] f(y,z)dG(y) .
\end{eqnarray*}
We only establish the first result as the other can be handled with the same techniques. By assumption (A.1) the function $w$ and its first-order partial derivatives are bounded on $(0,z]$ and one has :
\begin{eqnarray*}
|I_t-I| &\le&\sup_{y \le z}\Big|  s_t(y) 
{\partial \over \partial u}w \left[k(y)+t\pi(t)s_t(y),k(z)+t\zeta(t)s_t(z)\right]\\&&- s(y) {\partial \over \partial u}w \left[k(y),k(z)\right]  \Big | f(y,z)\int_0^z dG(y).
\end{eqnarray*}
Adding and substracting appropriate terms and observing that both $k$ and 
$s$ are bounded by 1, one has :
\begin{eqnarray*}
|I_t-I| 
&\le&
\sup_{y \le z}\Big| s_t(y)-s(y) \Big |\times  \sup_{y \le z} \Big \{
\Big|{\partial \over \partial u}w \left[k(y)+t\pi(t)s_t(y),k(z)+t\zeta(t)s_t(z)\right]\Big|f(y,z) \Big \}
\\&&+ \sup_{y \le z} \Big \{\Big |
{\partial \over \partial u}w \left[k(y)+t\pi(t)s_t(y),k(z)+t\zeta(t)s_t(z)\right]-{\partial \over \partial u}w \left[k(y),k(z)\right] \Big |f(y,z) \Big \}.
\end{eqnarray*}
The fact that $s_t \longrightarrow s$, as $t \rightarrow 0$ entails that $|s_t(y)-s(y) | \longrightarrow 0$, as $t \rightarrow 0$. Consequently, the first term in the right-hand side of the above inequality tends to 0, as $t$ tends to 0. The second term also tends to 0, as $t$ goes to 0. This is due to the continuity of $w$ and its first-order partial derivatives.  
%
%
It results from above that, $\phi$ is Hadamard-differentiable at $k \in \mathcal K$, tangentially to $\mathcal K'$, with      derivative $ \phi'_{w,f}[k]$, given for all $s \in \mathcal K$ by 
\begin{eqnarray*}
 \phi'_{w,f}[k](s)&=& 
\int_0^z \left\{ s(y) {\partial \over \partial u}w[k(y),k(z)] + s(z) {\partial \over \partial v}w[k(y),k(z)] \right \}f(y,z)dG(y).
\end{eqnarray*}
%
%
%
Since $\sqrt{n}[G_n - G]$ converge weakly to $ \mathbb B \circ G$, where $ \mathbb B$ stands for the standard Brownian bridge, it follows from the  functional delta method (see, e.g., \cite{vdv}) that $\sqrt{n}[\phi_{w,f}( G_n)-\phi_{w,f}(G)]=\sqrt{n}[Ph_{w,f,G_n}-Ph_{w,f,G}]=\mathbb{W}_n(w,f)$ converges in distribution to the Gaussian variable
\begin{eqnarray*} 
 \phi'_{w,f}[G](\mathbb B \circ G)&=& 
\int_0^z \left\{  \mathbb B \circ G (y) {\partial \over \partial u}w[G(y),G(z)] + \mathbb B \circ G(z) {\partial \over \partial v}w[G(y),G(z)] \right \}f(y,z)dG(y)\\
 &=: &\mathbb{W}(w,f).
\end{eqnarray*}
Since the class of functions $\mathcal{W\times\mathcal{F}}$ is shown to be Donsker according to Part $I$, we can infer that the process $\{\mathbb{W}_n(w,f): w\in\mathcal{W}, f\in\mathcal{F}\}$ converge in distribution to $\mathbb{W}(w,f)$ which is a zero-mean Gaussian process, with covariance kernel given, for all $(w,f)$ and $(\widetilde w, \widetilde f)$, by 
\begin{eqnarray*}
{\rm cov}(\mathbb{W}(w,f),\mathbb{W}(\widetilde w, \widetilde f))& = &
\int_0^{z}\int_0^{z} a_1(x,y)[G(x)\wedge G(y) - G(x)G(y)]dG(x)dG(y), \nonumber  \\
&& + [1-G(z)]\int_0^{z}\int_0^{z} a_2(x,y)G(y)dG(x)dG(y) \nonumber  \\
&& +[1- G(z)] \int_0^{z}\int_0^{z} a_3(x,y)G(x)dG(x)dG(y)  \nonumber  \\
&& + G(z)[1-G(z)] \int_0^{z}\int_0^{z} a_4(x,y)dG(x)dG(y), \nonumber 
\end{eqnarray*} 
where
\begin{eqnarray*}
a_1(x,y) & =& \frac{\partial w}{\partial u}[G(x),G(z)]f(x,z)\frac{\partial \widetilde w}{\partial u}[G(y),G(z)] \widetilde f(y,z)\\
a_2(x,y) & =& \frac{\partial w}{\partial
u}[G(x),G(z)]f(x,z)\frac{\partial \widetilde w}{\partial v}[G(y),G(z)] \widetilde f(y,z)\\
a_3(x,y)& =&\frac{\partial w}{\partial v}[G(x),G(z)]f(x,z)\frac{\partial \widetilde w}{\partial u}[G(y),G(z)] \widetilde f(y,z)\\
a_4(x,y)& =& \frac{\partial w}{\partial v}[G(x),G(z)]f(x,z)\frac{\partial \widetilde w}{\partial v}[G(y),G(z)] \widetilde f(y,z).
\end{eqnarray*}

\subsection{Part IV} 
Here we show that the couple of processes $(\mathbb{G}_n, \mathbb{W}_n)$  converges weakly to joint process $(\mathbb{G},\mathbb{W})$ which is a zero-mean Gaussian process. To this end, we show that it is tight and that its finite marginal distributions converge to those of a Gaussian process.

The tightness follows immediately  from Parts $I$ and $III$ where, it is proved that $\mathbb{G}_n$ converges weakly to a tight Gaussian process $\mathbb{G} \in l^{\infty}(\mathcal{H}_{z,G})$,  and $\mathbb{W}_n $ converges weakly to a tight Gaussian process $\mathbb{W} \in l^{\infty}(\mathcal{W\times F})$. 

For the study of the finite dimensional distributions, we have to show that for all $\alpha_1, \ldots, \alpha_m,$   $\beta_1, \ldots, \beta_{\ell} \in \mathbb R$ and $(w_l,f_l), (\widetilde w_i, \widetilde f_i) \in \mathcal {W \times F}$, $l=1, \ldots,m$, $i=1, \ldots, \ell$, the linear combination 
\begin{equation}\label{cl}
\sum_{l=1}^m \alpha_l\mathbb{G}_n(h_{w_l,f_l,G})+\sum_{i=1}^{\ell} \beta_i 
\mathbb{W}_n(\widetilde w_i, \widetilde f_i)
\end{equation}
is asymptotically Gaussian. For this, we make use of the asymptotic linearity of the two processes $\mathbb{G}_n$ and $\mathbb{W}_n$. For larger values of $n$, the latter can be expressed in terms of the former. Indeed for all $(w,f) \in \mathcal {W \times F}$ denote by  $L_{w,f}$ the Hadamard derivative of $\phi_{w,f}$ at $G$ ; that is $L_{w,f}=\phi'_{w,f}[G]$. Then for larger values of $n$ one has
$$
\mathbb{W}_n(w,f)= \sqrt{n}(Ph_{w,f,G_n}-Ph_{w,f,G})=
L_{w,f}(\sqrt{n}[G_n -G])+o_P(1).
$$
Since $G_n (\cdot)=n^{-1}\sum_{j=1}^n \mathbb{I}(Y_j\leq \cdot)=n^{-1}\sum_{j=1}^n \mathbb{I}_{[Y_j, \infty)}(\cdot)$, using the linearity of $L_{w,f}$, we obtain for $n$ large enough that 
\begin{eqnarray*}
\mathbb{W}_n (w,f)
& = & \frac{1}{\sqrt{n}}\sum_{j=1}^n \left[ L_{w,f}( \mathbb{I}_{[Y_j, \infty)}(\cdot) - L_{w,f}(G)\right] +o_P(1)\\
&= & \mathbb{G}_n\left( L_{w,f}(\mathbb{I}_{[Y_j, \infty)}(\cdot)\right) +o_P(1).  
\end{eqnarray*}
Combining this with the linearity of $\mathbb{G}_n$ we obtain, for all $\alpha_1, \ldots, \alpha_m,$ $ \beta_1, \ldots, \beta_{\ell} \in \mathbb R$ and $(w_l,f_l)$, $(\widetilde w_i, \widetilde f_i)$ $\in$ $ \mathcal {W \times F}$, $l=1, \ldots,m$, $i=1, \ldots, \ell$, that
\begin{eqnarray*}
&&\sum_{l=1}^m \alpha_l \mathbb{G}_n(h_{w_l,f_l,G})+\sum_{i=1}^{\ell} \beta_i 
\mathbb{W}_n(\widetilde w_i, \widetilde f_i)\\
&=&
\sum_{l=1}^m \alpha_l \mathbb{G}_n(h_{w_l,f_l,G}) + 
\sum_{i=1}^{\ell} \beta_i 
\mathbb{G}_n \left(L_{\widetilde w_i, \widetilde f_i}(\mathbb{I}_{[Y_j, \infty)}(\cdot)\right)+o_P(1)\\
&=&
\mathbb{G}_n \left(\sum_{l=1}^m \alpha_l h_{w_l,f_l,G} + 
\sum_{i=1}^{\ell} \beta_i L_{\widetilde w_i, \widetilde f_i}(\mathbb{I}_{[Y_j, \infty)}(\cdot)\right)+o_P(1).
\end{eqnarray*}
Recall that $\mathbb{G}_n$ is the empirical process and that the function $\sum_{l=1}^m \alpha_l h_{w_l,f_l,G} + 
\sum_{i=1}^{\ell} \beta_i L_{\widetilde w_i, \widetilde f_i}(\mathbb{I}_{[Y_j, \infty)}(\cdot)$ belongs to $L^2(P)$. 
Then it follows that the random variable defined in (\ref{cl})
is asymptotically Gaussian, and hence the finite marginal distributions of the process $(\mathbb{G}_n, \mathbb{W}_n)$, 
$$\left( \mathbb{G}_n(h_{w_1,f_1,G}),\cdots,\mathbb{G}_n(h_{w_m,f_m,G}),\mathbb{W}_n(\widetilde w_1, \widetilde f_1),\cdots,\mathbb{W}_n(\widetilde w_{\ell}, \widetilde f_{\ell}) \right)  $$  are asymptotically Gaussian too. Combining this with the tightness argument enable us to conclude that the joint process $(\mathbb{G}_n, \mathbb{W}_n)$ converges weakly to the process $(\mathbb{G}, \mathbb{W})$ which  is Gaussian and centered.

\end{document}